\documentclass[aip,jcp,reprint,groupedaddress]{revtex4-1}
\usepackage{graphicx,amssymb,amsmath}
\usepackage{epstopdf}
\begin{document}
\title{Nonlinear electro-osmosis in dilute non-adsorbing polymer solutions with low ionic strength}
\author{Yuki Uematsu}
\affiliation{Department of Physics, Kyoto University, Kyoto 606-8502, Japan}
\date{\today}
\begin{abstract}
Nonlinear behavior of electro-osmosis in dilute non-adsorbing polymer solutions with low salinity is investigated 
with Brownian dynamics simulations and a kinetic theory. 
In the Brownian simulations, 
the hydrodynamic interaction between the polymers and a no-slip wall 
is considered with Rotne-Prager approximation of Blake tensor. 
In a plug flow under a sufficiently strong applied electric field, the polymer migrates toward the bulk, forming a depletion layer thicker than the equilibrium one.
Consequently, the electro-osmotic mobility increases nonlinearly with the electric field and gets saturated.
This nonlinear mobility qualitatively 
does not depend on the details of rheological properties of the polymer solution.
Analytical calculation of the kinetic theory for the same system reproduces quantitatively well the results of the Brownian dynamics simulation.
\end{abstract}

\maketitle
\section{Introduction}

Electro-osmosis is observed widely in many systems such as colloids, porous materials and biomembranes. 
It characterizes the properties of interfaces between solids and electrolyte solutions.\cite{DukhinDerjaguin1974,RusselSavilleSchowalter1989}
Recently, there have been growing interests in applications of electro-osmosis.
For instance, it is used to pump fluids in microfluidic devices since it is more efficient than pressure-driven flow.\cite{SquiresQuake2005}
Application to an electrical power conversion is also very fascinating in chemical engineering.\cite{WangChengWangKiu2009,HeydenBonthuisSteinMeyerDekker2006}
When the electrokinetic properties of a surface are characterized by zeta potential, the Smoluchowski equation is often employed with measurements of the electro-osmotic or electrophoretic mobilities. 
However, one has to consider the validity of the mentioned equation more seriously.
It is derived from the Poisson-Boltzmann equation and the Newton's constitutive equation for viscous fluids. 
The zeta potential is defined as the electrostatic potential at the plane where a no-slip boundary condition is assumed.
When these equations are not validatable, the Smoluchowski equation is also questionable.  
In the cases of strong-coupling double layer,\cite{KimNetz2006} inhomogeneity of viscosity and dielectric constant near interface,\cite{UematsuAraki2013,BonthuisNetz2012} and non-Newtonian fluids\cite{ZhaoYang2010,ZhaoYang2011,ZhaoYang2011-2,ZhaoYang2013}, for example, the Poisson-Boltzmann and/or the simple hydrodynamic equations sometimes do not work well. 

In order to control the electrokinetic properties of charged capillaries, the structures of liquid interfaces contacted with charged surfaces are modified by grafting or adding polymers.\cite{ZnalezionaPeterKnobMaierSevcik2008}
In capillary electrophoresis, for example, the electro-osmotic flow is reduced by grafted polymers on the interfaces.
About surfaces with end-grafted charged and uncharged polymers, several studies have been also reported.\cite{Ohshima1994,Ohshima1995,HardenLongAjdari2001,HickeyHolmHardenSlater2011,RaafatniaHickeyHolm2014,RaafatniaHickeyHolm2015}
Under a weak applied electric field, the grafted polymer still remains in the equilibrium configuration and the resultant electro-osmotic velocity behaves linearly with respect to the electric fields. 
To measure the mobility of such a surface, hydrodynamic screening and anomalous charge distributions due to the grafted polymers are important.\cite{Ohshima1994,Ohshima1995,HardenLongAjdari2001,HickeyHolmHardenSlater2011,RaafatniaHickeyHolm2014,RaafatniaHickeyHolm2015}
When a sufficiently large electric field is applied, the polymers are deformed by the flow and electric field, and thus, the electro-osmotic velocity becomes nonlinear.\cite{HardenLongAjdari2001}
It should be noted that the end-grafted polymers cannot migrate toward the bulk since one of the ends is fixed on the surfaces. 

When we add polymers into solutions, a depletion or adsorption layer is often formed near a solid wall as well as diffusive layers of ions in equilibrium states.
The interaction between the polymers and the wall determines whether the polymers are depleted from or adhere to the surfaces.
The thickness of the depletion or adsorption layers is of the same order of the gyration length of polymers. 
When polymers adhere to the wall, the viscosity near the wall becomes large, so that the electro-osmosis mobility is much suppressed.\cite{HickeyHardenSlater2009}
Moreover, it is known that an adsorption layer of charged polymers can change the sign of the mobility.\cite{UematsuAraki2013,LiuEricksonLiKrull2004,DangerRamondaCottet2007,FengAdachiKobayashi2014,MarconiMonteferranteMelchionna2014}
The curvature of the surface also modulates the surface charge density, and even increases the mobility beyond the suppression caused by the viscosity enhancement.\cite{HickeyHardenSlater2009}

Electro-osmosis of a non-adsorbing polymer solution was analyzed by two length scales; the equilibrium depletion length $\delta_0$ and the Debye length $\lambda$.\cite{UematsuAraki2013,ChangTsao2007}
In the depletion layer, the viscosity is estimated approximately by that of the pure solvent and it is smaller than the solution viscosity in the bulk.
When the Debye length is smaller than the depletion length, 
the electro-osmotic mobility is larger than that estimated by the bulk value of viscosity.
Typically for 10\thinspace mM electrolyte solutions, one has $\lambda\approx$ 3\thinspace nm and $\delta_0\approx$ 100\thinspace nm. 
In such a case,\cite{BerliOlivares2008,OlivaresVeraCandiotiBerli2009,Berli2013} an electro-osmotic flow of high shear rate is localized in the distance $\lambda$ from the wall.
Thus, the electro-osmotic flow profile and resultant electro-osmotic mobility are almost independent of the polymers. 
Actually, such behaviors are experimentally observed in solutions of carboxymethyl cellulose with urea.\cite{OlivaresVeraCandiotiBerli2009}
On the other hand, in the solutions of small polymers with low salinity, typically for 0.1\thinspace mM electrolyte solutions $\lambda\approx$ 30\thinspace nm and $\delta_0\approx$ 5\thinspace nm, the electro-osmotic mobility is suppressed by the polymeric stress.\cite{ChangTsao2007}

When a sufficiently strong electric field is applied, the electro-osmosis of a polymer solution shows non-linear behaviors.\cite{BelloBesiRezzonicoRighettiCasirghi1994,OlivaresVeraCandiotiBerli2009}
These nonlinearities are theoretically analyzed by models of uniform non-Newtonian shear thinning fluids.\cite{ZhaoYang2010,ZhaoYang2011,ZhaoYang2011-2,ZhaoYang2013}
Assuming that polymers still remain localized in interfacial layers and the viscosity depends on the local shear rate as in power-law fluids, their phenomenological parameters are different from those in the bulk since the concentration in the interfacial layers is different from the bulk concentration.\cite{OlivaresVeraCandiotiBerli2009}
Thus, understanding of nonlinear electro-osmosis still remains phenomenological.
Furthermore, when shear flow is applied to polymer solutions near a wall, it is experimentally and theoretically confirmed that cross-stream migration is induced toward the bulk.\cite{JendrejackSchwartzPabloGraham,MaGraham2005,HernandezOrtizMaPabloGraham}
The concentration profiles of the polymer near the wall are calculated and the depletion length dynamically grows tenfold larger than the gyration radius.\cite{MaGraham2005}
However, these hydrodynamic effects in electrokinetics have not been studied so far to the best of our knowledge.

In this context, the present paper discusses another origin of nonlinearity which is induced by the hydrodynamic interaction between the polymer and wall.
Mainly the situation of $\delta_0\ll\lambda$ is concerned.
For this purpose, this paper is organized as follows.
Section \ref{sec2} presents a toy model for nonlinear electro-osmosis of dilute polymer solutions.
Section \ref{sec3} describes Brownian dynamics simulation.
Section \ref{sec4} presents results of the simulation.
Section \ref{sec5} discusses analytical approach for nonlinear electro-osmosis by using kinetic theory of cross-stream migration.\cite{MaGraham2005} 
Section \ref{sec6} outlines the main conclusions.
%The author hopes these nonlinear effects are utilized for electro-osmotic pumping and energy conversion.

\section{A Toy Model}
\label{sec2}
First, we propose a toy model for electro-osmosis of polymer solutions. 
A dilute solution of non-adsorbing short polymers is considered.
The viscosity of the solution is given by
\begin{equation}
\eta=\eta_0(1+\eta_\mathrm{sp}),
\end{equation}
where $\eta_0$ is the viscosity of the pure solvent, and $\eta_\mathrm{sp}$ is the specific viscosity of the solution.
The gyration length of the polymers is defined as $\delta_0$, which is of the same order of the equilibrium depletion length.
It is assumed that the short polymers have $\delta_0\approx 100\thinspace$nm.
Ions are also dissolved in the solution with the Debye length $\lambda$. 
When a well deionized water is considered, the Debye length is of the order of  $\lambda\approx 10^3\thinspace$nm although such a salt-free water is hardly realized owing spontaneous dissolutions of carbon dioxides.
The interfacial structure near a charged surface is characterized by $\lambda$ and $\delta_0$.
When an external electric field is applied, a shear flow is locally imposed within the distance $\lambda$ from the wall, and the resultant shear rate is 
\begin{equation}
\dot{\gamma}\approx\frac{\mu_0 E}{\lambda},
\label{eq2}
\end{equation}
where $\mu_0$ is the electro-osmotic mobility for the pure solvent and is estimated typically as $\mu_0\approx 10^{-8}\thinspace$m$^2$/(V$\cdot\thinspace$s).
According to the studies of the cross-stream migration in the uniform shear flow\cite{MaGraham2005}, the depletion layer thickness depends on the shear rate,
\begin{equation}
\delta\approx \delta_0 (\tau{\dot\gamma})^2,
\label{eq3}
\end{equation}
where $\tau$ is the characteristic relaxation time of the polymers,
\begin{equation}
\tau\approx\frac{\eta_0 {\delta_0}^3}{k_\mathrm{B}T}, 
\end{equation}
and is typically $10^{-4}\thinspace\mathrm{s}$.
Using eqs.(\ref{eq2}) and (\ref{eq3}), the depletion length in the presence of the applied electric field $E$ can be expressed by
\begin{equation}
\delta\approx\left\{\begin{array}{cl}
\delta_0 & \mathrm{for}\thickspace E<E_0,\\
&\\
\displaystyle\left(\frac{E}{E_0}\right)^2\delta_0&\mathrm{for}\thickspace  E_0\le E\le E_1,\\
&\\
\lambda&\mathrm{for}\thickspace  E_1<E,
\end{array}\right.
\end{equation}
where $E_0=\lambda/\tau\mu_0$, and $E_1=E_0\sqrt{\lambda/\delta_0}$.
Here, for simplicity, we assume that the depletion length does not exceed the Debye length.
The effective viscosity in the double layer is given by 
\begin{equation}
\eta_\mathrm{eff}\approx\eta_0\left[1+\eta_\mathrm{sp}\left(1-\frac{\delta}{\lambda}\right)\right],
\end{equation}
and the nonlinear mobility can be estimated by $\mu\approx \mu_0(\eta_0/\eta_\mathrm{eff})$.
Therefore, the mobility is obtained as
\begin{equation}
\mu\approx\left\{\begin{array}{cl}
\displaystyle \frac{\mu_0}{1+\eta_\mathrm{sp}(1-(\delta_0/\lambda))} & \mathrm{for}\thickspace E<E_0,\\
&\\
\displaystyle \frac{\mu_0}{1+\eta_\mathrm{sp}(1-(E/E_1)^2)}& \mathrm{for}\thickspace E_0\le E\le E_1,\\
&\\
\mu_0 &\mathrm{for}\thickspace  E_1<E.
\end{array}\right.
\end{equation}
\begin{figure}[t]
\centering
\includegraphics[width=0.48\textwidth]{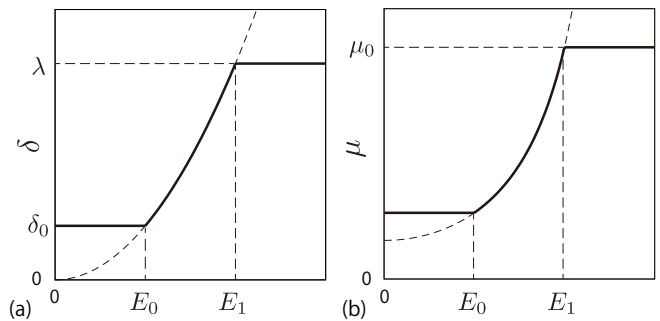}
\caption{
(a) The depletion length as a function of the applied electric field. 
(b) The electro-osmotic mobility as a function of the applied electric field.
}
\label{figure1}
\end{figure}
Fig.~\ref{figure1} (a) shows schematically the thickness of a depletion layer as a function of electric field strength. 
Fig.~\ref{figure1} (b) is the nonlinear electro-osmotic mobility. 
The mobility increases and is saturated with increasing electric filed.
The threshold electric field $E_0$ is typically $10^3\thinspace$V/m that is experimentally accessible.
\section{Model for simulation}
\label{sec3}
\begin{figure}[t]
\centering
\includegraphics[width=0.5\textwidth]{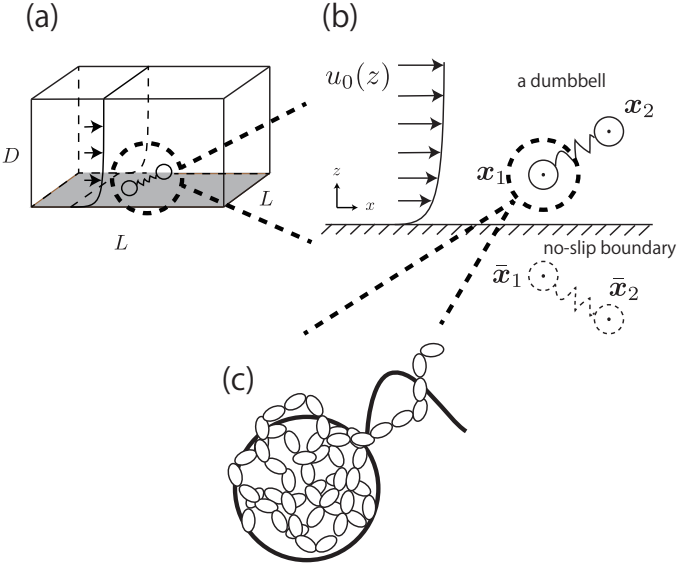}
\caption{
(a) A dumbbell in the simulated box with $L\times L\times D$. 
Periodic boundary condition has imposed at $x$-$y$ plane.
(b) A $x$-$z$ projection of the dumbbell in the electro-osmotic flow.
(c) An enlarged illustration of the bead in the dumbbell.
It is composed of a large number of monomeric units of the polymers.
}
\label{figure2}
\end{figure}

In this section, our method of Brownian dynamic simulation is described.
%A polymer chain in a dilute solution is modeled by a dumbbell. 
%The dumbbell is dissolved in an electrolyte solution, and a solid wall with the no-slip boundary condition at $z=0$ is considered.
As shown in Fig.~\ref{figure2}(a), a dumbbell is simulated in a electrolyte solution with a no-slip boundary at $z=0$.
The dumbbell shows a dilute solution behavior.
The solvent is described by a continuum fluid with the viscosity $\eta_0$ and fills up the upper half of space $(z>0)$.
Electrolytes are also treated implicitly with the Debye length $\lambda=\kappa^{-1}$.
The dumbbell has two beads whose hydrodynamic radii are $a$, and actually the bead consists of many monomeric units of the polymer (see Fig.~\ref{figure2}(c)).
The positions of the beads are represented by $\boldsymbol{x}_1$ and $\boldsymbol{x}_2$ (see Fig.~\ref{figure2}).
Then we solve overdamped Langevin equations\cite{ErmakMcCammon1978} given by
\begin{eqnarray}
\frac{dx_{n\alpha}}{dt}=u_0(z_n)\delta_{\alpha x}+\sum_{m,\beta}\left(\mathrm{G}^{nm}_{\alpha\beta}F_{m\beta}+k_\mathrm{B}T\nabla_{m\beta} \mathrm{G}^{nm}_{\alpha\beta}\right)+\xi_{n\alpha},&&\nonumber\\
\hspace{30mm}\textrm{for}\thickspace n=1,2,\thickspace\alpha=x,y,z,&& 
\label{eq:Langevin}
\end{eqnarray}
where $X_{n\alpha}$ is the $\alpha$ component of a vector $\boldsymbol{X}_n$. 
$u_0(z)$ is the external plug flow, $\nabla_{n\alpha}=\partial/\partial x_{n\alpha}$, $\mathbf{G}$ is the mobility tensor, $\boldsymbol{F}_n=-\nabla_n U$ is the force exerted on the $n$th bead, $U$ is the interaction energy given as a function of $\boldsymbol{x}_n$, and $k_\mathrm{B}T$ is the thermal energy. 
$\boldsymbol{\xi}_n$ is the thermal noise which satisfies the fluctuation-dissipation relation as
\begin{equation}
\langle \xi_{n\alpha}(t)\xi_{m\beta}(t')\rangle=2k_\mathrm{B}T\mathrm{G}^{nm}_{\alpha\beta}\delta(t-t').
\end{equation}
To include the effects of the no-slip boundary, Rotne-Prager approximation for Blake tensor\cite{Blake1971} is used for the mobility tensor for distinct particles $(n\neq m)$, \cite{WajnrybMizerskiZukSzymczak2013,ZukWajnrybMizerskiSzymczak2014} although it is valid only for particles separated far away. 
In this study, we neglect lubrication corrections for nearby particles.\cite{SwanBrady2007}
The Blake tensor for the velocity at $\boldsymbol{x}_2$ induced by a point force at $\boldsymbol{x}_1$ with the no-slip boundary at $z=0$ is given by the Oseen tensor and the coupling fluid-wall tensor as,\cite{Blake1971}
\begin{equation}
\mathrm{G}^\mathrm{B}_{\alpha\beta}(\boldsymbol{x}_2,\boldsymbol{x}_1)=\mathrm{S}_{\alpha\beta}(\boldsymbol{q})+\mathrm{G}^\mathrm{W}_{\alpha\beta}(\boldsymbol{x}_2,\boldsymbol{x}_1),
\label{eq10}
\end{equation}
where $\boldsymbol{q}=\boldsymbol{x}_2-\boldsymbol{x}_1$, $\boldsymbol{R}=\boldsymbol{x}_2-\bar{\boldsymbol{x}}_1$, and $\bar{\boldsymbol{x}}_1$ is the mirror image of $\boldsymbol{x}_1$ with respect to the plane $z=0$ (see Fig.~\ref{figure2}(b)).
The Oseen tensor given by
\begin{equation}
\mathrm{S}_{\alpha\beta}(\boldsymbol{q})=\frac{1}{8\pi\eta_0}\left(\frac{\delta_{\alpha\beta}}{q}+\frac{q_\alpha q_\beta}{q^3}\right),
\end{equation}
where $q$ is the magnitude of $\boldsymbol{q}$, and the second term in eq.~(\ref{eq10}) is
\begin{eqnarray}
\mathrm{G}^\mathrm{W}_{\alpha\beta}(\boldsymbol{x}_2,\boldsymbol{x}_1)&=&-\mathrm{S}_{\alpha\beta}(\boldsymbol{R})+z_1^2(1-2\delta_{\beta z})\nabla^2_R\mathrm{S}_{\alpha\beta}(\boldsymbol{R})\nonumber\\
&&-2z_1(1-2\delta_{\beta z})\mathrm{S}_{\alpha z,\beta}(\boldsymbol{R}),
\end{eqnarray}
where 
\begin{equation}
\mathrm{S}_{\alpha\beta,\gamma}(\boldsymbol{q})=\nabla_{q\gamma}\mathrm{S}_{\alpha\beta}(\boldsymbol{q}),
\end{equation}
and $\nabla_{q\gamma}=\partial/\partial q_{\gamma}$.
The Rotne-Prager approximation of the Blake tensor is given by\cite{KimNetz2006,WajnrybMizerskiZukSzymczak2013,ZukWajnrybMizerskiSzymczak2014,HansenHinczewskiNetz2011,SwanBrady2007}
\begin{equation}
\mathrm{G}^\mathrm{RPB}_{\alpha\beta}(\boldsymbol{x}_2,\boldsymbol{x}_1)=\left\{
\begin{array}{l}
\displaystyle\left(1+\frac{a^2}{6}\nabla_1^2+\frac{a^2}{6}\nabla_2^2\right)\mathrm{G}^\mathrm{B}_{\alpha\beta}(\boldsymbol{x}_2,\boldsymbol{x}_1)+\mathcal{O}(a^4)\\
\hspace{45mm}\mathrm{for}\thickspace q>2a,\\
\\
\displaystyle\frac{1}{6\pi\eta_0 a}\left[\delta_{\alpha\beta}-\frac{9q}{32a}\left(\delta_{\alpha\beta}-\frac{q_\alpha q_\beta}{3q^2}\right)\right]\vspace{2mm}\\
\hspace{2mm}\displaystyle+\left(1+\frac{a^2}{6}\nabla_1^2+\frac{a^2}{6}\nabla_2^2\right)\mathrm{G}^\mathrm{W}_{\alpha\beta}(\boldsymbol{x}_2,\boldsymbol{x}_1)+\mathcal{O}(a^4)\\
\hspace{45mm}\mathrm{for}\thickspace q\le 2a.
\end{array}
\right.
\end{equation}
The mobility tensor for the self part is given by\cite{KimNetz2006,WajnrybMizerskiZukSzymczak2013,ZukWajnrybMizerskiSzymczak2014,HansenHinczewskiNetz2011,SwanBrady2007}
\begin{eqnarray}
\mathrm{G}^\mathrm{self}_{\alpha\beta}(z)&=&\lim_{\boldsymbol{x}\to\boldsymbol{x}_1}\mathrm{G}^\mathrm{RPB}_{\alpha\beta}(\boldsymbol{x},\boldsymbol{x}_1)\nonumber\\
&=&\left(
\begin{array}{ccc}
\mu_\parallel(z)&0&0\\
0&\mu_\parallel(z)&0\\
0&0&\mu_\perp(z)
\end{array}\right),
\end{eqnarray}
where 
\begin{eqnarray}
\mu_\parallel(z)&=&\frac{1}{6\pi\eta_0 a}\left[1-\frac{9a}{16z}+\frac{1}{8}\left(\frac{a}{z}\right)^3\right]+\mathcal{O}(a^4),\\
\mu_\perp(z)&=&\frac{1}{6\pi\eta_0 a}\left[1-\frac{9a}{8z}+\frac{1}{2}\left(\frac{a}{z}\right)^3\right]+\mathcal{O}(a^4).
\end{eqnarray}
Finally we obtain the mobility tensor as
\begin{equation}
\mathrm{G}^{nm}_{\alpha\beta}=\delta_{nm}\mathrm{G}^\mathrm{self}_{\alpha\beta}(z_n)+(1-\delta_{nm})\mathrm{G}^\mathrm{RPB}_{\alpha\beta}(\boldsymbol{x}_n,\boldsymbol{x}_m).
\end{equation}
The non-uniform mobility term in eq.~(\ref{eq:Langevin}) can be simplified within using the Rotne-Prager approximation of the Blake tensor because the relation
\begin{equation}
\sum_{\beta=x,y,z}\nabla_{m\beta} \mathrm{G}^\mathrm{RPB}_{\alpha\beta}(\boldsymbol{x}_n,\boldsymbol{x}_m)=0,
\end{equation}
is hold.
Thus, the non-uniform mobility term is rewritten by
\begin{equation}
\sum_{m,\beta}\nabla_{m\beta}\mathrm{G}^{nm}_{\alpha\beta}=\delta_{\alpha z}\nabla_{nz}\mu_\perp(z_n).
\end{equation}

The interaction energy includes spring and bead-wall interaction given by
\begin{equation}
U=U^\mathrm{s}(q)+\sum_{n=1,2}U^\mathrm{w}(z_n),
\end{equation}
where $U^\mathrm{s}$ is the spring energy as 
\begin{equation}
U^\mathrm{s}(q)=\left\{\begin{array}{cc}
\displaystyle\frac{H}{2}q^2,& \textrm{Hookian dumbbells},\\
&\\
\displaystyle -\frac{H}{2}q_0^2\ln\left[1-\left(\frac{q}{q_0}\right)^2\right],& \textrm{FENE dumbbells},
\end{array}\right.
\end{equation}
where a FENE dumbbell stands for a finitely extensible nonlinear elastic dumbbell, and a parameter $b=Hq_0^2/k_\mathrm{B}T$ is defined for convenience.
$U^\mathrm{w}$ is the bead-wall interaction,\cite{BitsanisHadziioannou1990} which is purely repulsive as
\begin{equation}
U^\mathrm{w}(z)=\left\{\begin{array}{lc}
\displaystyle w\left[\frac{2}{5}\left(\frac{a}{z}\right)^{10}-\left(\frac{a}{z}\right)^4+\frac{3}{5}\right] &\mathrm{for}\thickspace z\le a,\\
\\
0&\mathrm{for}\thickspace z>a.
\end{array}\right.
\end{equation}
Eq.~(\ref{eq:Langevin}) is numerically solved.
Reflection boundary condition is set at $z=D$.
When the center of the dumbbell goes across the boundary, the $z$-coordinate of each beads are transformed from $z$ to $2D-z$.
For the lateral directions, the periodic boundary conditions are imposed. 
The size of the lateral directions is $L\times L$.

\section{Results of simulation}
\label{sec4}
The concentration and velocity profiles are calculated by
\begin{equation}
c(z)=\frac{1}{L^2}\left\langle\delta\left(z-\frac{z_1+z_2}{2}\right)\right\rangle,
\end{equation}
and 
\begin{equation}
\delta u(z)=\frac{1}{\eta_0 L^2}\left\langle\sum_{n=1,2}\textrm{min}(z,z_n)F_{nx}\right\rangle,
\label{eq.deltau}
\end{equation}
where $\delta(z)$ is the delta function, $\delta u(z)=u(z)-u_0(z)$ is the velocity increment due to the polymeric stress, and $\langle\cdots\rangle$ means a statistical average in steady states.
The derivation of eq.~(\ref{eq.deltau}) is written in Appendix A.

The imposed electro-osmotic flow $u_0(z)$ is given by
\begin{equation}
u_0(z)=\mu_0 E \left (1-\mathrm{e}^{-\kappa z}\right),
\end{equation}
where $\mu_0$ is the electro-osmotic mobility in the pure solvent, and $E$ is the applied electric field.\cite{RusselSavilleSchowalter1989}
Eq.~(\ref{eq:Langevin}) is rewritten in a dimensionless form with the length scale $\delta_0=\sqrt{k_\mathrm{B}T/H}$ and time scale $\tau=6\pi\eta_0 a/4H$.
The different types of dumbbells are simulated with the parameters noted in Table~\ref{table1}.
\begin{table}
\caption{
Simulation parameters.
$N_\mathrm{t}$ is the number of total steps, $N_\mathrm{i}$ is the number of interval steps for observation, $N_\mathrm{m}$ is the number of the sampling for each parameter set, and $\Delta\tau$ is the time increment.
}
\label{table1}
\begin{center}
\begin{tabular}{cccc}
\hline
&Hookian&FENE $b=600$& FENE $b=50$\\\hline
$N_\mathrm{t}$&$5\times 10^{10}$& $5\times 10^{10}$& $25\times 10^{10}$\\
$N_\mathrm{i}$&$5\times 10^{3}$& $5\times 10^{3}$& $25\times 10^{3}$\\
$N_\mathrm{m}$&3&3&5\\
$\Delta\tau$&0.01&0.0025 & 0.0001\\
\hline
\end{tabular}
\end{center}
\end{table}
It should be noted that the simulated systems are treated as dilute systems and the linearity with respect to the bulk polymer concentration is preserved.
After sample averaging, we obtain the concentration at the upper boundary $c(D)$, which slightly deviates from $(L^2D)^{-1}$ because of the inhomogeneity near the surface.
Hereafter, we define the normalized concentration as,
\begin{equation}
C(z)=\frac{c(z)}{c(D)}.
\end{equation}
As well as the concentration profile, the velocity increment $\delta u(z)$ has the linearity with respect to $c(D)$. 
For convenience, we set a characteristic concentration $c_\mathrm{b}=0.1{\delta_0}^{-3}$, and the nonlinear electro-osmotic mobility is defined by
\begin{equation}
\mu(E)=\mu_0+\frac{c_\mathrm{b}}{c(D)}\frac{\delta u(D)}{E}.
\end{equation}
The top boundary is placed at $D=100\delta_0$, the lateral size is set to $L=1000\delta_0$, and the Debye length is set to $\lambda=\kappa^{-1}=10\delta_0$.
We also set $w=3k_\mathrm{B}T$, and a hydrodynamic parameter $h^*$ as\cite{MaGraham2005}
\begin{equation}
h^*=\frac{a}{\sqrt{\pi}\delta_0}=0.25.
\end{equation}

\begin{figure}
\centering
\includegraphics[width=0.48\textwidth]{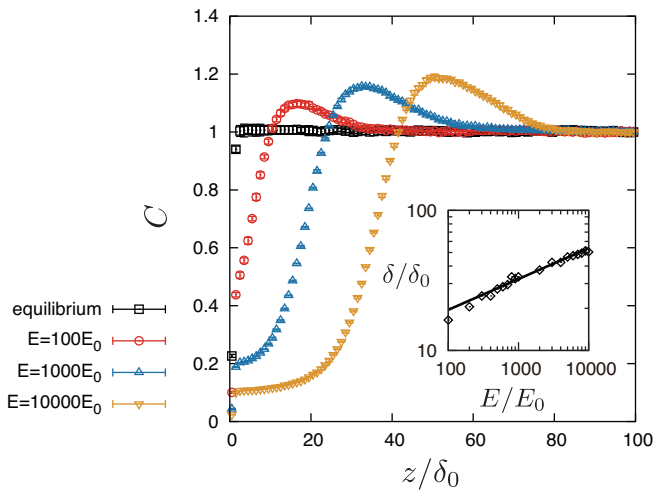}
\caption{
The concentration profiles of the Hookian dumbbells with varied applied electric fields.
The inset shows the depletion length as a function of the applied field.
The points are obtained by the Brownian dynamics simulation and the line is fitted by $\delta/\delta_0=A(E/E_0)^B$, where $A=7.08$ and $B=0.22$.
}
\label{figure3}
\end{figure}

Fig.~\ref{figure3} shows the steady state profiles of the Hookian-dumbbell concentration as functions of the distance from the wall. 
In the equilibrium state of $E=0$, the profile shows the depletion layer whose width is of the same order as the gyration length $\delta_0$.
When the applied electric field is increased stronger, the depletion layer becomes larger than the equilibrium one and a peak is formed.
The inset in Fig.~\ref{figure3} shows the depletion length as a function of the applied field.
The depletion length is defined by the position of the concentration peak.
It shows a power-law behavior and its exponent is 0.22, which is much smaller than $2.0$ in the case of a uniform shear flow.\cite{MaGraham2005}
The value of the concentration at the peak also increases as the electric field is enlarged.

\begin{figure}
\centering
\includegraphics[width=0.48\textwidth]{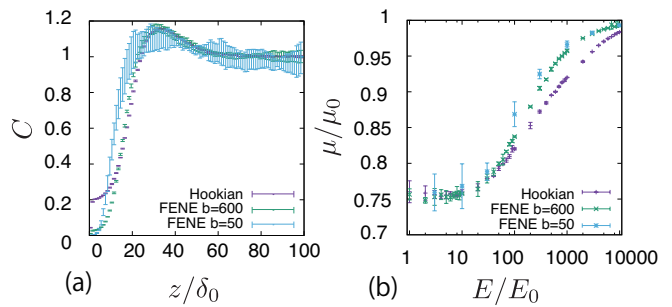}
\caption{
(a) The concentration profiles of the polymers at $E=1000E_0$ in the different types of dumbbells.
(b) Plots of the nonlinear electro-osmotic mobilities with respect to $E$.
}
\label{figure4}
\end{figure}

The results mentioned above are for the Hookian dumbbell which is infinitely extensible with the shear deformation.
To consider more realistic polymers, the finitely extensible nonlinear elastic dumbbell is simulated.
Fig.~\ref{figure4} (a) shows the concentration profiles at $E=1000E_0$.
Interestingly, the one-peak behaviors are also observed in the FENE dumbbells. In the case of the Hookian dumbbell, the concentration near the surface remains finite.
On the other hand, in the case of the FENE dumbbells, the concentrations near the surface are negligibly small.
Fig.~\ref{figure4} (b) plots the electro-osmotic mobilities with respect to the applied electric field. It is clearly shown the resultant electro-osmosis grows  nonlinearly with respect to applied electric field.
When the applied field gets stronger, the mobility increases and is saturated similarly to that in the toy model.
The two types of the dumbbells have different rheological properties from each other at the bulk,\cite{ChristiansenBird1977,BirdDotsonJohnson1980,BirdCurtissArmstrongHassager1987}
 so that this nonlinearity is not owing to the rheological properties of the dumbbells.
On the other hand, the mobility is almost constant for $E\lesssim 10E_0$, and this threshold of the linearity is larger than $E_0$, that is predicted by the toy model.
Likewise the saturation is observed when $E\approx 10^4E_0$, which is larger than $E_1$.

\begin{figure}
\centering
\includegraphics[width=0.48\textwidth]{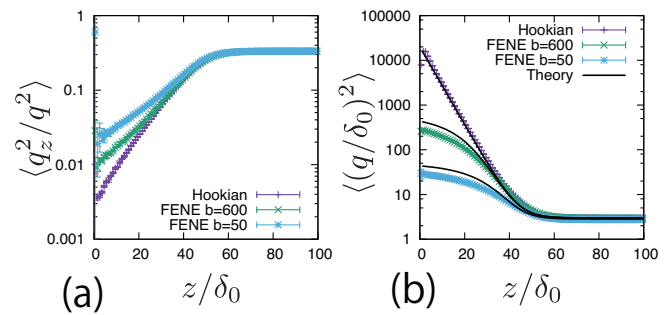}
\caption{
(a) The profiles of $\langle q_z^2/q^2\rangle$ at $E=1000E_0$ in the different types of the dumbbells.
(b) The profiles of $\langle (q/\delta_0)^2\rangle$ at $E=1000E_0$ in the different types of the dumbbells.
The curved lines are calculated by eqs.~(\ref{qq1}) and (\ref{qq2}).
}
\label{figure5}
\end{figure}

To clarify the difference of the profiles near the surface, $\langle q^2\rangle$ and $\langle q_z^2/q^2\rangle$ are plotted with respect to the distance from the surface.
Fig.~\ref{figure5} (a) shows the profiles of $\langle q_z^2/q^2\rangle$.
In the bulk, they approach to $1/3$, which means the dumbbells are distributed isotropically.
Near the surface, the polymers are inclined by the shear flow.
Concerning the angles between the $z$-axis and the dumbbell direction, that of the Hookian dumbbell is the largest among them.
Fig.~\ref{figure5} (b) plots the profiles of $\langle q^2\rangle$. 
In the bulk, they approach to $3\delta_0$ which is the equilibrium value of them.
Near the surface, they become larger since the polymers are elongated by the shear flow.
In the case of FENE dumbbells, the saturations of the dumbbell length are observed. 
These behaviors are largely different from the minor difference in the concentration profiles.

\section{Kinetic theory}
\label{sec5}
In this section, a kinetic theory for a dumbbell is developed based on Ma-Graham theory\cite{MaGraham2005}.
The probability function $\Psi(\boldsymbol{x}_1,\boldsymbol{x}_2,t)$ obeys the continuity equation
\begin{equation}
\frac{\partial\Psi}{\partial t}=-\nabla_1\cdot(\dot{\boldsymbol{x}}_1\Psi)-\nabla_2\cdot(\dot{\boldsymbol{x}}_2\Psi),
\end{equation}
where $\dot{\boldsymbol{x}}_n$ is the flux velocity being given by\cite{DoiEdwards1986}
\begin{equation}
\dot{x}_{n\alpha}=u_0(z_n)\delta_{x\alpha}-\sum_{m,\beta}\mathrm{G}^{nm}_{\alpha\beta}\nabla_{m\beta}(U+k_\mathrm{B}T\ln\Psi).
\end{equation}
In the kinetic model, the beads are treated as point-like particles.
Thus the mobility tensor is obtained by using $\mathbf{G}^\mathrm{B}$ instead of $\mathbf{G}^\mathrm{RPB}$ for both the self and distinct parts.
The continuity equation can be rewritten with $\boldsymbol{q}$ and $\boldsymbol{r}$ as 
\begin{equation}
\frac{\partial\Psi}{\partial t}=-\nabla_r\cdot(\dot{\boldsymbol{r}}\Psi)-\nabla_q\cdot(\dot{\boldsymbol{q}}\Psi),
\end{equation}
where
$\boldsymbol{r}=(\boldsymbol{x}_1+\boldsymbol{x}_2)/2$ is the center of the mass of the dumbbell. 
We also define $\nabla_1$ and $\nabla_2$ as
\begin{eqnarray}
\nabla_{1}&=&\frac{1}{2}\nabla_{r}-\nabla_{q},\\
\nabla_{2}&=&\frac{1}{2}\nabla_{r}+\nabla_{q}.
\end{eqnarray}
Then, the probability function is also regarded as a function of $\boldsymbol{r}$ and $\boldsymbol{q}$.
Here we neglect the interaction between the wall and beads. 
The flux velocities for $\boldsymbol{r}$ and $\boldsymbol{q}$ are obtained by
\begin{eqnarray}
\dot{r}_\alpha&=&\frac{1}{2}[u_0(z_1)+u_0(z_2)]\delta_{x\alpha}+\frac{1}{2}\bar{\mathrm{G}}_{\alpha\beta}F^\mathrm{s}_\beta\nonumber\\
&&+\frac{k_\mathrm{B}T}{2}\bar{\mathrm{G}}_{\alpha\beta}\nabla_{q,\beta}\ln\Psi-\mathrm{D}^\mathrm{K}_{\alpha\beta}\nabla_{r,\beta}\ln\Psi,\\
\label{rflux}
\dot{q}_\alpha&=&[u_0(z_2)-u_0(z_1)]\delta_{x\alpha}-\hat{\mathrm{G}}_{\alpha\beta}F^\mathrm{s}_\beta\nonumber\\
&&+\frac{k_\mathrm{B}T}{2}\bar{\mathrm{G}}_{\alpha\beta}\nabla_{r,\beta}\ln\Psi-k_\mathrm{B}T\hat{\mathrm{G}}_{\alpha\beta}\nabla_{q,\beta}\ln\Psi,
\end{eqnarray}
where $\boldsymbol{F}^\mathrm{s}=-\nabla_{1}U^\mathrm{s}$ is the spring force, and $\mathbf{D}^\mathrm{K}$ is the Kirkwood diffusion tensor which characterizes the diffusivity of macromolecules, given by
\begin{equation}
\mathbf{D}^\mathrm{K}=\frac{k_\mathrm{B}T(\mathbf{G}^{11}+\mathbf{G}^{12}+\mathbf{G}^{21}+\mathbf{G}^{22})}{4}.
\end{equation} 
$\bar{\mathbf{G}}$ and $\hat{\mathbf{G}}$ are a variation of the mobility tensors defined by
\begin{eqnarray}
\bar{\mathbf{G}}&=&\mathbf{G}^{11}-\mathbf{G}^{12}+\mathbf{G}^{21}-\mathbf{G}^{22},\\
\hat{\mathbf{G}}&=&\mathbf{G}^{11}-\mathbf{G}^{12}-\mathbf{G}^{21}+\mathbf{G}^{22}.
\end{eqnarray}
The concentration field $c(\boldsymbol{r},t)$ can be obtained by integrating the probability function with respect to the spring coordinate. 
It is given by
\begin{equation}
c(\boldsymbol{r},t)=\int \Psi(\boldsymbol{r},\boldsymbol{q},t) d\boldsymbol{q}.
\end{equation}
We also define the probability function only for the spring coordinate as 
\begin{equation}
\hat\Psi(\boldsymbol{q},t;\boldsymbol{r})=\frac{\Psi(\boldsymbol{r},\boldsymbol{q},t)}{c(\boldsymbol{r},t)}.
\end{equation}
These new fields satisfy the continuity conditions, such that
\begin{eqnarray}
\frac{\partial c}{\partial t}&=&-\nabla_{r}\cdot(c\langle\dot{\boldsymbol{r}}\rangle_q),\\
\frac{\partial\hat\Psi}{\partial t}&=&-\nabla_{q}\cdot(\hat\Psi \dot{\boldsymbol{q}}),
\label{continuityq}
\end{eqnarray}
where $\langle\cdots\rangle_q$ means the average with the spring coordinate, as
\begin{equation}
\langle\cdots\rangle_q=\frac{\int\cdots\Psi(\boldsymbol{r},\boldsymbol{q},t)d\boldsymbol{q}}{c(\boldsymbol{r},t)}=\int\cdots\hat{\Psi}(\boldsymbol{r},\boldsymbol{q},t)d\boldsymbol{q}.
\end{equation}
For the limit of $q\ll r$, $\bar{\mathbf{G}}$ and $\mathbf{D}^\mathrm{K}$ can be expanded with $\boldsymbol{r}$.
With keeping only the leading term, we obtain
\begin{eqnarray}
\bar{\mathbf{G}}&=&\frac{3}{32\pi\eta_0 z^2}\left(\begin{array}{ccc}
-q_z&0&-\chi q_x\\
0&-q_z&-\chi q_y\\
\chi q_x&\chi q_y&-2q_z
\end{array}\right)+\cdots,\\
&&\nonumber\\
\mathbf{D}^\mathrm{K}&=&\frac{k_\mathrm{B}T}{12\pi\eta_0 a}\left[\mathbf{I}+\frac{3a}{4}\mathbf{S}(\boldsymbol{q})\right]+\cdots,
\end{eqnarray}
where 
\begin{equation}
\chi=\left[1+\frac{q_x^2+q_y^2}{4z^2}\right]^{-5/2}.
\end{equation}
It should be noted that the approximation is more accurate than that in a previous study\cite{MaGraham2005} since they considered only $\chi\approx 1$, which is not satisfied near the surface.
With the approximation, eq.(\ref{rflux}) is averaged by $\hat\Psi$, and
finally we obtain the concentration flux for $z$ direction as
\begin{equation}
c\langle\dot{r}_z\rangle_q=cu_\mathrm{mig}(z)-\frac{d}{dz}\left[c\langle \mathrm{D}^\mathrm{K}_{zz}\rangle_q\right],
\label{flux}
\end{equation}
where
\begin{eqnarray}
u_\mathrm{mig}(z)&=&\frac{1}{2}\langle\bar{\mathrm{G}}_{z\beta}F^\mathrm{s}_\beta-k_\mathrm{B}T\nabla_{q\beta}\bar{\mathrm{G}}_{z\beta}\rangle_q\nonumber\\
&=&\frac{3}{64\pi\eta_0 z^2}\nonumber\\
&&\times\left\langle\chi(q_xF_x^\mathrm{s}+q_yF_y^\mathrm{s})-2q_zF_z^\mathrm{s}-2k_\mathrm{B}T(\chi-1)\right\rangle_q.\nonumber\\
\end{eqnarray}
Eq.~(\ref{flux}) indicates two opposite fluxes of the polymers due to the external flow field.
One is the migration flux from the wall toward the bulk and originates from the hydrodynamic interaction between the wall and the force dipoles.\cite{MaGraham2005}
The other is the diffusion flux from the bulk to the surface wall and is not found in the case of polymers in uniform shear flows.\cite{MaGraham2005}
It should be noted that the second flux includes not only the ordinary diffusion flux $\langle D_{zz}^\mathrm{K}\rangle_q\nabla_{r,z}c$, but also the diffusion flux due to the $q$-inhomogeneity, $c\nabla_{r,z}\langle D_{zz}^\mathrm{K}\rangle_q$.
When the external shear flow is uniform, the second flux vanishes, and the depletion length is proportional to the square of the shear rate since the migration velocity is proportional to the normal stress difference.\cite{MaGraham2005}
In the case of a plug flow, the diffusion flux suppresses the growth of the depletion layer and it may answer why the exponent of the depletion length is much smaller than 2.0 in the uniform shear flow.
In steady states of the electro-osmosis, the total flux in eq.~(\ref{flux}) becomes zero, and thus,
\begin{equation}
\frac{dc}{dz}=\frac{c}{\langle D^\mathrm{K}_{zz}\rangle_q}\left(u_\mathrm{mig}-\frac{d\langle D_{zz}^\mathrm{K}\rangle_q}{dz}\right).
\end{equation}
This equation shows the migration flux and the diffusion flux are balanced at the peak of the concentration profiles.
Finally the concentration profile is calculated by
\begin{equation}
c=c_\mathrm{b}\exp\left[\int^z_D\frac{1}{\langle D_{zz}^\mathrm{K}\rangle_q}\left(u_\mathrm{mig}(z')-\frac{d\langle D_{zz}^\mathrm{K}\rangle_q}{dz}\right)dz'\right].
\end{equation}
The resultant flow profile can be calculated by
\begin{eqnarray}
\delta u(z)&=&-\frac{1}{\eta_0}\int^z_0\sigma_{xz}^\mathrm{p}(z')dz',
\end{eqnarray}
where $\boldsymbol{\sigma}^\mathrm{p}$ is the polymeric part of the stress tensor as
\begin{equation}
\boldsymbol{\sigma}^\mathrm{p}=c\langle\boldsymbol{q}\boldsymbol{F}^\mathrm{s}\rangle_q-ck_\mathrm{B}T\mathbf{I}.
\end{equation}

To obtain explicit expressions of $c$ and $\delta u$, it it necessary to estimate $u_\mathrm{mig}$, $\langle D_{zz}^\mathrm{K}\rangle_q$, and $\boldsymbol{\sigma}^\mathrm{p}$.
For this purpose, eq.~(\ref{continuityq}) should be analyzed.
However, eq.~(\ref{continuityq}) is highly complicated.
Even without the wall effects, it cannot be solved exactly, so that several approximation methods have been proposed.\cite{ZylkaOttinger1989} 
For simplicity, all the hydrodynamic interactions are ignored, and thus, the continuity equation is given
\begin{equation}
\frac{\partial\hat{\Psi}}{\partial t}=-\nabla_{q\alpha}\left[\left(\frac{du_0}{dz}q_z\delta_{z\alpha}-\frac{2F^\mathrm{s}_\alpha}{6\pi\eta_0 a}\right)\hat{\Psi}-\frac{2k_\mathrm{B}T}{6\pi\eta_0 a}\nabla_{q\alpha}\hat{\Psi}\right].
\label{dynamicsq}
\end{equation}
For the Hookian dumbbell eq.~(\ref{dynamicsq}) can be solved for the second moment of $\boldsymbol{q}$, and for the FENE dumbbell pre-averaged approximation\cite{BirdDotsonJohnson1980,BirdCurtissArmstrongHassager1987}
 is employed.
The curved lines in Fig.~\ref{figure5} are calculated with these approximations, and they agree quantitatively well with the simulation results.
In Appendix B, approximated expressions for these quantities of the Hookian and FENE dumbbells are written. 

\begin{figure}
\centering
\includegraphics[width=0.48\textwidth]{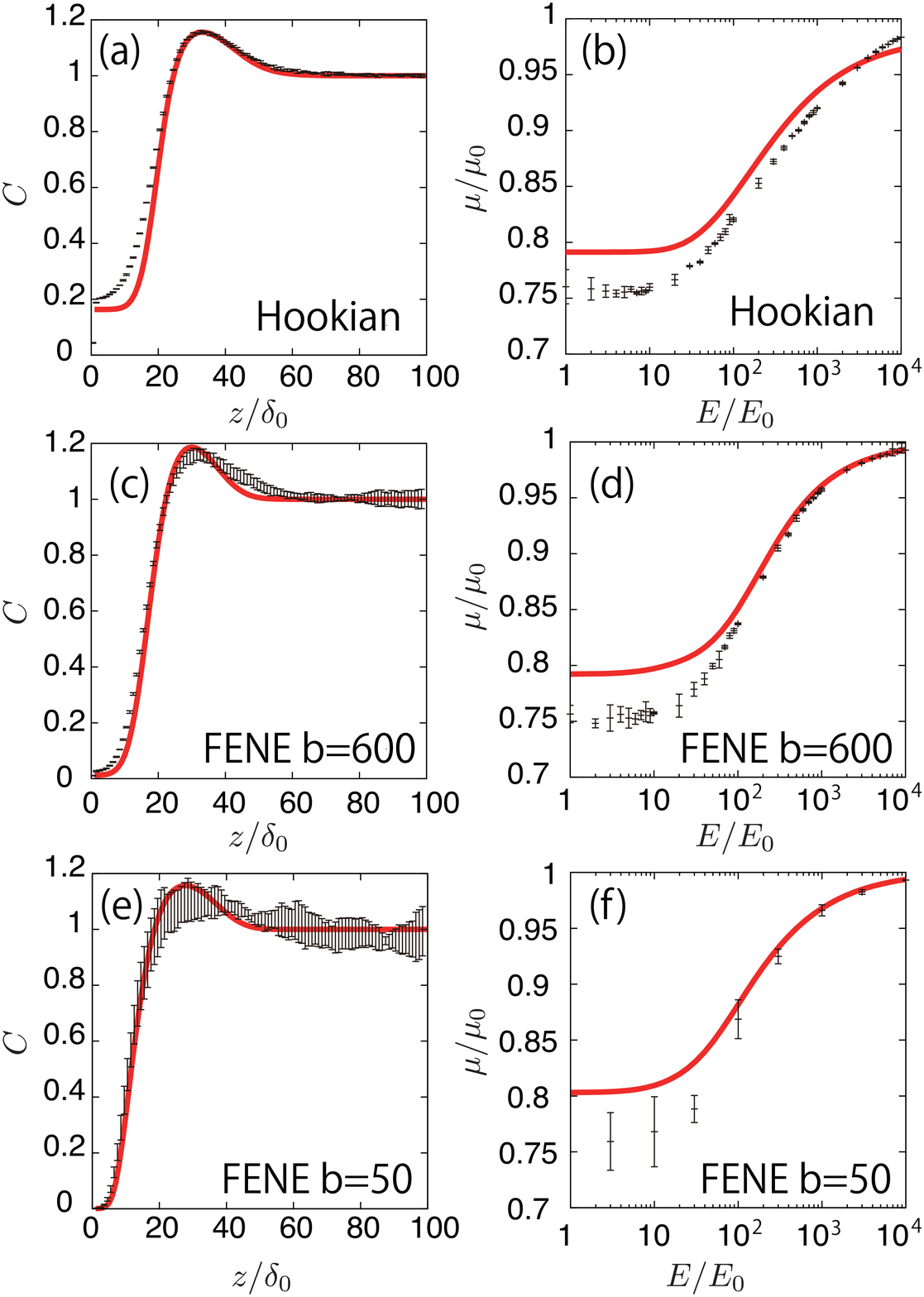}
\caption{
(a) The concentration profiles of the Hookian dumbbell as a function of the distance from the surface.
The points show the simulation results and the curved line is calculated by the kinetic theory.
(b) The nonlinear electro-osmotic mobilities of the Hookian dumbbell as a function of a applied electric field.
(c) and (d) Those for the FENE dumbbell with $b=600$.
(e) and (f) Those for the FENE dumbbell with $b=50$.
}
\label{figure6}
\end{figure}
Fig.~\ref{figure6} (a), (c) and (e) show the concentration profiles for the applied field $E=1000E_0$.
The points are obtained by the Brownian dynamics simulation and the curved lines are obtained by the kinetic theory. 
The theoretical calculations quantitatively cover well the simulations.
Moreover, they reproduce the differences in the concentration near the surface between the Hookian and FENE dumbbells, since the migration velocities can be approximately proportional to $\langle\chi\rangle_q$ (see Appendix \ref{app2}), and it is much suppressed in the case of the Hookian dumbbells.
Fig.~\ref{figure6} (b), (d), and (f) show the nonlinear electro-osmotic mobilities with respect to the applied field.
The theoretical curved lines also have an acceptable tendency with the simulation results. 
However, they are not so consistent with the simulation results in weak applied electric fields since the equilibrium depletion layer is not considered in the kinetic theory. 

\section{Summary and remarks}
\label{sec6}
With Brownian dynamics simulations, nonlinear behaviors of electro-osmosis of dilute polymer solutions are studied. The simulation results agree with a toy-model and analytical calculations of a kinetic theory. The main results are summarized below.
\begin{itemize}
\item[(i)] Under an external plug flow, the polymer migrates toward the bulk. The concentration profile of the polymer shows a depletion layer and a single peak. 
The thickness of the depletion layer depends on the electric field. 
At the peak, the migration flux is balanced to the diffusion flux.
\item[(ii)] The growth of depletion layer leads to increment and saturation in the electro-osmotic mobility. 
Qualitatively this behavior does not depend on the rheological properties of the dumbbells. 
\item[(iii)] Analytical calculation of the concentration and the nonlinear mobility by the kinetic theory is in agreement with the Brownian dynamics simulation.
The threshold of the electric field for the nonlinear growth and saturation of the mobility is much larger than the prediction of the toy model, since the diffusive flux suppresses the migration toward the bulk due to the inhomogeneous shear flow. 

\end{itemize}
We conclude this study with the following remarks.
\begin{itemize}
\item[(1)]
Nonlinear electro-osmosis with $\lambda\ll\delta_0$ has already been observed experimentally.\cite{BerliOlivares2008,OlivaresVeraCandiotiBerli2009} 
They reported the mobility is increased with increasing the electric field.
However, the nonlinear electro-osmosis with $\lambda\gg\delta_0$ has not been reported experimentally, and therefore, experimental verification of our findings is highly desired.
\item[(2)]
It would be a future problem whether the hydrodynamic interaction between the polymers and the surface plays an important role in electro-osmosis in polymer solutions even though $\lambda\ll\delta_0$ or not.
In this case the elongation of the polymers is strongly inhomogeneous under the plug flow with a short Debye length, and thus more realistic chain models should be considered.
\item[(3)]
Addition of charged polymers into solutions can change the direction of the linear electro-osmotic flow.\cite{UematsuAraki2013,FengAdachiKobayashi2014}
When a sufficiently strong electric field is applied to this system, the direction of the flow might recover its original one.
It needs to be investigated theoretically and experimentally.
\end{itemize}
\section*{Acknowledgement}
The author is grateful to O. A. Hickey, T. Sugimoto, R. R. Netz, and M. Radtke for helpful discussions.
The author also thanks M. Itami, M. R. Mozaffari, and T. Araki for their critical reading of the manuscript.
This work was supported by the JSPS Core-to-Core Program ``Nonequilibrium dynamics of soft matter and information'', a Grand-in-Aid for JSPS fellowship, and KAKENHI Grant No. 25000002.
\appendix
\section{Derivation of the velocity equation for Brownian dynamics simulation}
In this appendix, the derivation of eq.~(\ref{eq.deltau}) is explained.
The velocity field induced by the polymer is given by
\begin{equation}
\delta u(z)=-\frac{1}{\eta_0}\int^z_0\sigma^\mathrm{p}_{xz}(z)dz,
\end{equation}
and the polymeric part of the stress tensor is obtained by averaging those of the microscopic expression in the lateral directions as
\begin{equation}
\sigma^\mathrm{p}_{\alpha\beta}=\frac{1}{L^2}\int dxdy \hat{\sigma}^\mathrm{p}_{\alpha\beta}(\boldsymbol{x}).
\end{equation}
Here the microscopic expression of the stress tensor is given by
\begin{equation}
\hat{\sigma}^\mathrm{p}_{\alpha\beta}(\boldsymbol{x})=-\frac{1}{2}\sum_{n}\sum_{m\neq n}F_{nm,\alpha} x_{nm,\beta}\delta^\mathrm{s}_{nm}(\boldsymbol{x}),
\end{equation}
where $\boldsymbol{F}_{nm}$ is the force exerted on the $n$-th bead from the $m$-th bead and $\delta^\mathrm{s}_{nm}(\boldsymbol{x})$ is the symmetrized delta function given by
\begin{equation}
\delta_{nm}^\mathrm{s}(\boldsymbol{x})=\int^1_0 ds\delta(\boldsymbol{x}-s\boldsymbol{x}_n-(1-s)\boldsymbol{x}_m).
\end{equation}
The symmetrized delta function is integrated in the lateral directions as
\begin{eqnarray}
\bar{\delta}_{nm}^\mathrm{s}(z)&=&\int dx dy \delta_{nm}^\mathrm{s}(\boldsymbol{x})=\int^1_0 ds\delta(z-s\boldsymbol{z}_n-(1-s)\boldsymbol{z}_m)\nonumber\\
&=&\frac{\theta(z-z_m)-\theta(z-z_n)}{z_n-z_m},
\end{eqnarray}
where $\theta(z_n-z)=1-\theta(z-z_n)$.
Then we obtain 
\begin{eqnarray}
\int^z_0 dz'\bar{\delta}^{nm}_\mathrm{S}(z') &=&\frac{(z-z_n)\theta(z-z_n)-(z-z_m)\theta(z-z_m)}{z_m-z_n}\nonumber\\
&=&\frac{\min(z,z_n)-\min(z,z_m)}{z_n-z_m},
\end{eqnarray}
where $\min(z,z_n)=z\theta(z)-(z-z_n)\theta(z-z_n)$.
Finally, the velocity increment is expressed by
\begin{eqnarray}
\delta u(z)&=&\frac{1}{2\eta_0 L^2}\sum_{n,m}F_{nm,1}(z_n-z_m)\int^z_0dz'\bar\delta_{nm}^\mathrm{s}(z')\nonumber\\
&=&\frac{1}{2\eta_0 L^2}\sum_{nm}F_{nm,1}[\min(z,z_n)-\min(z,z_m)]\nonumber\\
&=&\frac{1}{\eta_0 L^2}\sum_n\min(z,z_n)F_{n,1}.
\end{eqnarray}
\section{Approximated expressions for kinetic theory}
\label{app2}
\subsection{a Hookian dumbbell}
Eq.~(\ref{dynamicsq}) can be rewritten in a closed form for the second moment of the spring coordinates in a steady state with an imposed plug flow. 
The solution is given by\cite{BirdCurtissArmstrongHassager1987}

\begin{equation}
\langle \boldsymbol{q}\boldsymbol{q}\rangle_q=\frac{k_\mathrm{B}T}{H}\left(\begin{array}{ccc}
1+2\phi^2&0&\phi\\
0&1&0\\
\phi&0&1
\end{array}\right),
\label{qq1}
\end{equation}
where 
\begin{equation}
\phi=\tau\frac{du_0}{dz}=\tau\kappa\mu_0 E e^{-\kappa z}.
\end{equation}
Therefore, we have
\begin{equation}
\langle\boldsymbol{qF}^\mathrm{s}\rangle_q=H\langle\boldsymbol{qq}\rangle_q,
\end{equation}
and the polymeric stress tensor is
\begin{eqnarray}
\boldsymbol{\sigma}^\mathrm{p}&=&c\langle \boldsymbol{q}\boldsymbol{F}^\mathrm{s}\rangle-ck_\mathrm{B}T\mathbf{I}\nonumber\\
&=&ck_\mathrm{B}T\left(\begin{array}{ccc}
2\phi^2&0&\phi\\
0&0&0\\
\phi&0&0
\end{array}\right).
\end{eqnarray}
The Kirkwood diffusion constant can be estimated by
\begin{eqnarray}
\langle\mathrm{D}^\mathrm{K}_{zz}\rangle_q&=&\frac{k_\mathrm{B}T}{12\pi\eta_0 a}\left[1+\frac{3a}{4}\left\langle\frac{1}{q}\left(1+\frac{q_z^2}{q^2}\right)\right\rangle\right]\nonumber\\
&\approx&\frac{k_\mathrm{B}T}{12\pi\eta_0 a}\left[1+\frac{3a}{4}\frac{\langle q^2+q_z^2\rangle}{\langle q^2\rangle^{3/2}}\right],
\end{eqnarray}
where the second term is split into the second order moments, and thus, we obtain
\begin{equation}
\langle D_{zz}^\mathrm{K}\rangle_q=\frac{k_\mathrm{B}T}{12\pi\eta_0 a}\left[1+\frac{3a}{4\delta}\frac{2(\phi^2+2)}{(2\phi^2+3)^{3/2}}\right].
\end{equation}
It is differentiated with $z$ as
\begin{equation}
\frac{d}{dz}\langle D_{zz}^\mathrm{K}\rangle_q=\frac{k_\mathrm{B}T}{12\pi\eta_0 a}\frac{3a}{4\delta}\frac{4\kappa\phi^2(\phi^2+3)}{(2\phi^2+3)^{5/2}}.
\end{equation}
The migration velocity can be estimated using the splitting approximation of the averages as
\begin{eqnarray}
u_\mathrm{mig}(z)&=&\frac{3k_\mathrm{B}T}{64\pi\eta_0 z^2}\left\langle\chi(q_x^2+q_y^2)-2\chi\right\rangle_q\nonumber\\
&\approx&\frac{3k_\mathrm{B}T}{32\pi\eta_0 z^2}\langle\chi\rangle_q\langle q_x^2+q_y^2-2\rangle_q\nonumber\\
&=&\frac{3k_\mathrm{B}T}{32\pi\eta_0 z^2}\langle\chi\rangle_q\phi^2,
\end{eqnarray}
where
\begin{eqnarray}
\langle\chi\rangle_q&=&\left\langle \left(1+\frac{q_x^2+q_y^2}{4z^2}\right)^{-5/2}\right\rangle_q\nonumber\\
&\approx&\left(1+\frac{\phi^2+1}{2z^2}\right)^{-5/2}.
\end{eqnarray}
\subsection{a FENE dumbbell}
The second moment of the spring coordinate for a FENE dumbbell can be obtained by pre-averaged closures of p-FENE model.\cite{BirdDotsonJohnson1980,BirdCurtissArmstrongHassager1987}
It is given by
\begin{equation}
\langle \boldsymbol{qq}\rangle_q=\frac{k_\mathrm{B}T}{H}\frac{\psi}{\phi}\left(\begin{array}{ccc}
1+2\psi^2&0&\psi\\
0&1&0\\
\psi&0&1
\end{array}\right),
\label{qq2}
\end{equation}
and
\begin{equation}
\langle \boldsymbol{qF}\rangle_q=k_\mathrm{B}T\left(\begin{array}{ccc}
1+2\psi^2&0&\psi\\
0&1&0\\
\psi&0&1
\end{array}\right),
\end{equation}
where 
\begin{equation}
\psi=6\sqrt{\frac{3+b}{54}}\sinh\left\{\frac{1}{3}\mathrm{arcsinh}\left[\frac{b\phi}{108}\left(\frac{3+b}{54}\right)^{-3/2}\right]\right\}.
\end{equation}
The polymer stress tensor is
\begin{eqnarray}
\boldsymbol{\sigma}^\mathrm{p}&=&ck_\mathrm{B}T\left(\begin{array}{ccc}
2\psi^2&0&\psi\\
0&0&0\\
\psi&0&0
\end{array}\right).
\end{eqnarray}
The Kirkwood diffusion constant is 
\begin{equation}
\langle D^\mathrm{K}_{zz}\rangle_q=\frac{k_\mathrm{B}T}{12\pi\eta_0 a}\left[1+\frac{3a}{4\delta}\sqrt{\frac{\phi}{\psi}}\frac{2(\psi^2+2)}{(2\psi^2+3)^{3/2}}\right],
\end{equation}
and its derivative is
\begin{eqnarray}
&&\frac{d}{dz}\langle D_{zz}^\mathrm{K}\rangle_q=\frac{k_\mathrm{B}T}{12\pi\eta_0 a}\frac{3a}{4\delta}\times \kappa\sqrt{\frac{\phi}{\psi}}\nonumber\\
&&\times\left[\phi\frac{d\psi}{d\phi}\frac{4\psi(\psi^2+3)}{(2\psi^2+3)^{5/2}}+\left(\frac{\phi}{\psi}\frac{d\psi}{d\phi}-1\right)\frac{2(\psi^2+2)}{(2\psi^2+3)^{3/2}}\right],\nonumber\\
\end{eqnarray}
where
\begin{eqnarray}
\frac{d\psi}{d\phi}&=&2\sqrt{\frac{b+3}{54}}\cosh\left\{\frac{1}{3}\mathrm{arcsinh}\left[\frac{b\phi}{108}\left(\frac{3+b}{54}\right)^{-{3/2}}\right]\right\}\nonumber\\
&&\times\frac{b}{108}\left[\left(\frac{b\phi}{108}\right)^2+\left(\frac{b+3}{54}\right)^3\right]^{-1/2}.
\end{eqnarray}
Finally the migration velocity is obtained as 
\begin{equation}
u_\mathrm{mig}\approx\frac{3k_\mathrm{B}T}{32\pi\eta_0 z^2}\langle\chi\rangle_q\psi^2,
\end{equation}
where
\begin{equation}
\langle\chi\rangle_q\approx\left(1+\frac{\psi}{\phi}\frac{\psi^2+1}{2z^2}\right)^{-5/2}.
\end{equation}

\end{document}